\documentstyle[12pt]{article}
\begin{document}
\centerline{\Large\bf Hamiltonian Formulation of the Pilot-Wave Theory}
\vskip .7in
\centerline{Dan N. Vollick}
\vskip .2in
\centerline{Irving K. Barber School of Arts and Sciences}
\centerline{University of British Columbia Okanagan}
\centerline{3333 University Way}
\centerline{Kelowna, B.C.}
\centerline{Canada}
\centerline{V1V 1V7}
\vskip 0.5in
\centerline{\bf\large Abstract}
\vskip 0.5in
\noindent
In the pilot-wave theory of quantum mechanics particles have definite positions and velocities and the system evolves deterministically. The velocity of a particle is determined by the wave function of the system (the guidance equation) and the wave function evolves according to Schrodinger's equation.

In this paper I first construct a Hamiltonian that gives Schrodinger's equation and the guidance equation for the particle. I then find the Hamiltonian for a relativistic particle in Dirac's theory and for a quantum scalar field.
\newpage
\section{Introduction}
In the standard approach to quantum mechanics the wave function provides a complete description of any system and is used to calculate probability distributions for observables associated with the system. In the pilot-wave theory pioneered by de Broglie \cite{de0,de1} in   the 1920's, particles have definite positions and velocities and are ``guided" by the wave function, which satisfies Schrodinger's equation. A similar approach was developed by Bohm \cite{Bohm1,Bohm1b} in the early 1950's (see \cite{Bohm5} and \cite{Hol2}  for extensive reviews).

For a single particle in the non-relativistic theory the velocity of the particle is given by
\begin{equation}
\frac{d\vec{X}}{dt}=\vec{J}(\vec{X},t)\;,
\end{equation}
where $\vec{X}$ is the particle's position and
\begin{equation}
\vec{J}=\frac{\hbar}{2im|\psi|^2}\left[\psi^*\vec{\nabla}\psi-\psi\vec{\nabla}\psi^*\right]\;.
\end{equation}
Particle trajectories are, therefore, integral curves to the vector field $\vec{J}$.

In this paper I first consider the pilot-wave theory for a non-relativistic particle in a potential. I construct a Hamiltonian that gives Schrodinger's equation and the guidance equation for the particle. This involves imposing a constraint and the use of Dirac's method of dealing with constrained dynamical systems \cite {Dir1, Dir2}. I then find the Hamiltonian for a relativistic particle in Dirac's theory and for a quantum scalar field.

A Hamiltonian formulation of the pilot-wave theory has also been developed by Holland \cite{Holland}. His approach is significantly different from the approach taken in this paper and will be discussed at the end of section 2.
\section{Non-Relativistic Pilot-Wave Theory}
In this section I construct a Hamiltonian that gives Schrodinger's equation and the guidance equation for a single particle and for a collection of particles.

The Lagrangian
\begin{equation}
L=\frac{1}{2}i\hbar\left[\psi^*\frac{\partial\psi}{\partial t}-\psi\frac{\partial\psi^*}{\partial t}\right]-\frac{\hbar^2}{2m}
\vec{\nabla}\psi\cdot\vec{\nabla}\psi^*-V\psi^*\psi
\end{equation}
gives Schrodinger's equation
\begin{equation}
i\hbar\frac{\partial \psi}{\partial t}=-\frac{\hbar^2}{2m}\nabla^2\psi+V\psi
\end{equation}
under a variation with respect to $\psi^*$ and its complex conjugate under a variation with respect to $\psi$.

The canonical momenta $\Pi_{\psi}=\partial L/\partial \dot{\psi}$ and $\Pi_{\psi^*}=\partial L/\partial \dot{\psi}^*$ are given by
\begin{equation}
\Pi_{\psi}=\frac{1}{2}i\hbar\psi^*\hskip 0.4in and \hskip 0.4in \Pi_{\psi^*}=-\frac{1}{2}i\hbar\psi\;.
\end{equation}
We therefore have the primary constraints
\begin{equation}
\phi_1=\Pi_{\psi}-\frac{1}{2}i\hbar\psi^*\approx 0\hskip 0.4in and \hskip 0.4in \phi_2=\Pi_{\psi^*}+\frac{1}{2}i\hbar\psi\approx 0\;,
\label{Pi}
\end{equation}
where $\approx$ denotes a weak equality, which can only be imposed after the Poisson brackets have been evaluated. These constraints satisfy
\begin{equation}
\{\phi_1(\vec{x}),\phi_1(\vec{y})\}=\{\phi_2(\vec{x}),\phi_2(\vec{y})\}=0
\end{equation}
and
\begin{equation}
\{\phi_1(\vec{x}),\phi_2(\vec{y})\}=-i\hbar\delta^{3}(\vec{x}-\vec{y})\;.
\end{equation}
These constraints are, therefore, second class constraints.

The canonical Hamiltonian density $h_C=\Pi_{\psi}\dot{\psi}+\Pi_{\psi^*}\dot{\psi}^*-L$ is given by
\begin{equation}
h_C=\frac{\hbar^2}{2m}\vec{\nabla}\psi\cdot\vec{\nabla}\psi^*+V\psi^*\psi+\phi_1\dot{\psi}+\phi_2\dot{\psi}^*
\end{equation}
and the total Hamiltonian density is given by
\begin{equation}
h_T=\frac{\hbar^2}{2m}\vec{\nabla}\psi\cdot\vec{\nabla}\psi^*+V\psi^*\psi+u_1\phi_1+u_2\phi_2\;,
\label{HT}
\end{equation}
where $u_1$ and $u_2$ are undetermined parameters.

For consistency we require that
\begin{equation}
\dot{\phi_1}=\{\phi_1,H_T\}\approx 0\hskip 0.4in and \hskip 0.4in \dot{\phi}_2=\{\phi_2,H_T\}\approx 0\;,
\end{equation}
where $H_T=\int h_Td^3x$ is the total Hamiltonian. These two equations give
\begin{equation}
u_1=\frac{i}{\hbar}\left[\frac{\hbar^2}{2m}\nabla^2\psi-V\psi\right]
\end{equation}
and
\begin{equation}
u_2=-\frac{i}{\hbar}\left[\frac{\hbar^2}{2m}\nabla^2\psi^*-V\psi^*\right]\;.
\end{equation}
Substituting these expressions for $u_1$ and $u_2$ into $H_T$ and integrating by parts gives
\begin{equation}
H_T=\frac{i}{\hbar}\int\left[\Pi_{\psi}\left(\frac{\hbar^2}{2m}\nabla^2\psi-V\psi\right)-
\Pi_{\psi^*}\left(\frac{\hbar^2}{2m}\nabla^2\psi^*-V\psi^*\right)\right]d^3x\;.
\end{equation}
The equation of motion for $\psi$ is
\begin{equation}
\dot{\psi}=\{\psi,H_T\}=\frac{i}{\hbar}\left(\frac{\hbar^2}{2m}\nabla^2\psi-V\psi\right)\;,
\end{equation}
which is Schrodinger's equation. The equation of motion for $\Pi_{\psi}$ is
\begin{equation}
\dot{\Pi}_{\psi}=\{\Pi_{\psi},H_T\}=\frac{i}{\hbar}\left(\frac{\hbar^2}{2m}\nabla^2\Pi_{\psi}-V\Pi_{\psi}\right)\;.
\end{equation}
From (\ref{Pi}) we see that this is Schrodinger's equation for $\psi^*$. Similar results hold for $\dot{\psi}^*$ and for
$\dot{\Pi}_{\psi^*}$. The constraints $\phi_1$ and $\phi_2$ can be promoted to strong equations if the Poisson bracket is replaced by the Dirac bracket.

A similar Hamiltonian density $h=-\frac{i\hbar}{2m}\vec{\nabla}\Pi\cdot\vec{\nabla}\psi -\frac{i}{\hbar}V\Pi\psi$ appears in Schiff
\cite{Sch1}, but he does not follow Dirac's procedure to obtain $h$. Gergely \cite{Ger1} follows Dirac's approach and finds the Hamiltonian density $h=\frac{\hbar^2}{2m}\vec{\nabla}\psi^*\cdot\vec{\nabla}\psi +V\psi^*\psi+\dot{\psi}\phi_1+\dot{\psi}^*\phi_2$. Schrodinger's equation and its complex conjugate then follow from the conditions $\dot{\phi}_1\simeq 0$ and $\dot{\phi}_2\simeq 0$
(see \cite{Str1} for a discussion on the Hamiltonian formulation of the DKP equation, which is also a first order equation, using Dirac's procedure).

In the pilot-wave theory the wave function is written as
\begin{equation}
\psi=Re^{iS/\hbar}\;,
\end{equation}
where $R$ and $S$ are real functions. The velocity of the particle is taken to be \cite{de1,Bohm1}
\begin{equation}
\frac{d\vec{X}}{dt}=\frac{1}{m}\vec{\nabla}S(\vec{X},t)\;,
\label{eom}
\end{equation}
where $\vec{X}(t)$ is the particle's position. The velocity can also be written as
\begin{equation}
\frac{d\vec{X}}{dt}=\frac{\vec{j}}{\psi^*\psi}=\frac{\hbar}{2im|\psi|^2}\left[\psi^*\vec{\nabla}\psi-\psi\vec{\nabla}\psi^*\right]\;,
\label{J}
\end{equation}
where $\vec{j}$ is the probability current density.

This equation of motion follows from the Hamiltonian
\begin{equation}
H_p=\frac{\vec{p}}{m}\cdot\vec{\nabla}S(\vec{X},t)\;.
\end{equation}
To see this consider the equations of motion
\begin{equation}
\dot{X}^k=\{X^k,H_p\}=\frac{1}{m}\left[\frac{\partial S}{\partial x^k}\right]_{\vec{x}=\vec{X}}
\label{x}
\end{equation}
and
\begin{equation}
\dot{p}_k=\{p_k,H_p\}=-\frac{p_l}{m}\left[\frac{\partial^2 S}{\partial x^k\partial x^l}\right]_{\vec{x}=\vec{X}}\;.
\label{p}
\end{equation}
Equation (\ref{x}) is the correct equation of motion for the particle. Equation (\ref{p}) gives the equation of motion for $p_k$, but $p_k$ is not related to the particle velocity in this theory. We can therefore solve for the particle's trajectory using (\ref{x}) alone and ignore (\ref{p}).

Now consider the Hamiltonian for the field and particle
\begin{equation}
H=H_T+H_P\;.
\label{Ham1}
\end{equation}
The equations of motion $\dot{\psi}=\{\psi,H\}$, $\dot{X}^k=\{X^k,H\}$ and $\dot{p}_k=\{p_k,H\}$ are the same as above. However the equation of motion for $\Pi_{\psi}$ is altered by the addition of $H_p$ to $H_T$:
\begin{equation}
\dot{\Pi}_{\psi}=\{\Pi_{\psi},H\}=\{\Pi_{\psi},H_T\}+\frac{p_k}{m}\{\Pi_{\psi},\partial_k S\}\;.
\label{Poisson}
\end{equation}
To evaluate $\{\Pi_{\psi},\nabla^kS\}$ it is convenient to write $\partial_kS$ as
\begin{equation}
\partial_kS(\vec{X},t)=\frac{\hbar}{2i}\int\left[\frac{\partial_k\psi(\vec{x},t)}{\psi(\vec{x},t)}-
\frac{\partial_k\psi^*(\vec{x},t)}{\psi^*(\vec{x},t)}\right]\delta^3(\vec{x}-\vec{X})d^3x\;.
\end{equation}
The last term in (\ref{Poisson}) given by
\begin{equation}
\frac{p_k}{m}\{\Pi_{\psi}(\vec{x}),\partial_kS(\vec{X},t)\}=\frac{p_k}{m\psi(\vec{x},t)}\frac{\partial}{\partial x^k}\delta^3(\vec{x}-\vec{X}).
\end{equation}
This term can be eliminated by imposing the constraint $p_k\approx 0$.
For consistency we require that
$\dot{p}_k=\{p_k,H\}\approx 0$. From equation (\ref{p}) we see that this is satisfied, so that no new constraints arise.
The Hamiltonian $H$ plus the constraint $p_k\approx 0$ therefore generates Schrodinger's equation and the equation of motion for the particle.

The generalization of (\ref{Ham1}) to a system of $N$ particles is straightforward. The Hamiltonian is given by
\begin{equation}
H=\frac{i}{\hbar}\int\left[\Pi_{\psi}\left(\sum_{k=1}^N\frac{\hbar^2}{2m_k}\nabla^2_k\psi-V\psi\right)-
\Pi_{\psi^*}\left(\sum_{k=1}^N\frac{\hbar^2}{2m_k}\nabla^2_k\psi^*-V\psi^*\right)\right]d^3x_1...d^3x_N
+\sum_{k=1}^N\frac{\vec{p}_k}{m_k}\cdot\vec{\nabla}_kS\;,
\end{equation}
where $\vec{\nabla}_kS$ can also be written as
\begin{equation}
\vec{\nabla}_kS=\sum_{k=1}^N\frac{\hbar}{2i|\psi|^2}\left[\psi^*\vec{\nabla}_k\psi-\psi\vec{\nabla}_k\psi^*\right]\;,
\end{equation}
$\psi=\psi(\vec{x}_1...,\vec{x}_N)$ and $V=V(\vec{x}_1...,\vec{x}_N,t)$. This Hamiltonian plus the constraints $\vec{p}_k\approx 0$ gives both the field and particle equations of motion.

A Hamiltonian formulation of the pilot-wave theory has also been developed by Holland \cite{Holland}. In his approach the wave function is written as $\psi=\sqrt{\rho}e^{iS/\hbar}$ and the Hamiltonian is given by
\begin{equation}
H_{tot}=H+\int\left\{-\pi_{\rho}\left[\frac{1}{m}\frac{\partial}{\partial q_i'}\left(\rho\frac{\partial S}{\partial q_i'}\right)\right]-\pi_{S}\left[\frac{1}{2m}\left(\frac{\partial S}{\partial q_i'}\right)\left(\frac{\partial S}{\partial q_i'}\right)+Q+V\right]\right\}d^3q',
\end{equation}
where
\begin{equation}
H=\sum_i\frac{p_i^2}{2m}+V+Q\;,
\end{equation}
$p_i$ is the momentum of the $i^{th}$ particle, V is the classical potential and $Q$ is the quantum potential which is given by
\begin{equation}
Q=-\frac{\hbar^2}{2m\sqrt{\rho}}\frac{\partial^2\sqrt{\rho}}{\partial q_i^2}\;.
\end{equation}
Hamilton's equations for $\dot{S}$ and $\dot{\rho}$ give Schrodinger's equation and Hamilton's equations for the particle degrees of freedom are given by
\begin{equation}
\dot{q}_i=p_i/m \;\;\;\;\;\;\;\;\;\;\;\;\; and \;\;\;\;\;\;\;\;\;\;\;\;\;  \dot{p}_i=-\frac{\partial}{\partial q_i}[V+Q]\;.
\end{equation}
Combining these two equations gives
\begin{equation}
m\ddot{q}_i=-\frac{\partial}{\partial q_i}[V+Q]\;,
\end{equation}
which is the second order form of the theory. This equation can be derived from the first order guidance equation, but it is not equivalent to the guidance equation. Holland then considers Liouville's equation and shows, for pure states, that the momentum is constrained to satisfy the guidance equation. This approach differs significantly from the approach taken in this paper.
\section{Relativistic Pilot-Wave Theory}
In this section I construct a Hamiltonian that gives the Dirac equation plus the guidance equation.

The Lagrangian for the Dirac field coupled to the electromagnetic field is given by
\begin{equation}
L=\frac{i}{2}\left[\bar{\psi}\gamma^{\mu}D_{\mu}\psi-(D^*_{\mu}\bar{\psi)}\gamma^{\mu}\psi\right]-m\bar{\psi}\psi\;,
\end{equation}
where $D_{\mu}=\partial_{\mu}+ieA_{\mu}$ and I have taken $\hbar=c=1$. There are two primary constraints
\begin{equation}
\phi_1=\Pi_{\psi}-\frac{1}{2}i\bar{\psi}\gamma^0\approx 0\hskip 0.4in and \hskip 0.4in \phi_2=\Pi_{\bar{\psi}}+\frac{1}{2}i\gamma^0\psi\approx 0
\label{const}
\end{equation}
that satisfy
\begin{equation}
\{\phi_{1_k}(\vec{x}),\phi_{1_{\ell}}(\vec{y})\}=\{\phi_{2_k}(\vec{x}),\phi_{2_{\ell}}(\vec{y})\}=0
\end{equation}
and
\begin{equation}
\{\phi_{1_k}(\vec{x}),\phi_{2_{\ell}}(\vec{y})\}=-i(\gamma^{0})_{\ell k}\delta^{3}(\vec{x}-\vec{y})\;.
\end{equation}
The constraints are, therefore, second class. Following the same procedure that was used for Schrodinger's equation gives
\begin{equation}
H_T=i\int\left\{\Pi_{\psi}\gamma^0\left[i\gamma^k\partial_k\psi-eA_{\mu}\gamma^{\mu}\psi-m\psi\right]+
\left[i(\partial_k\bar{\psi})\gamma^k+e\bar{\psi}A_{\mu}\gamma^{\mu}+m\bar{\psi}\right]\gamma^0\Pi_{\bar{\psi}}\right\}d^3x.
\end{equation}
In the pilot-wave theory the velocity of the particle is given by \cite{Bohm2,Hol1}
\begin{equation}
\frac{dX^{\mu}}{d\tau}=J^{\mu}
\end{equation}
where
\begin{equation}
J^{\mu}=\frac{\bar{\psi}\gamma^{\mu}\psi}{\sqrt{a^2+b^2}}\;,\;\;\;\;\;\;\;\;\;\; J^{\mu}J_{\mu}=1\;,
\end{equation}
\begin{equation}
a=\bar{\psi}\psi\;\;\;\;\;\;\;\; and \;\;\;\;\;\;\; b=i\bar{\psi}\gamma^5\psi\;.
\end{equation}
The particle Hamiltonian is
\begin{equation}
H_P=p_{\mu}J^{\mu}
\end{equation}
and the Hamiltonian for the field and particle is given by
\begin{equation}
H=H_T+H_p\;.
\end{equation}
As in the non-relativistic approach we must impose $p_{\mu}\approx 0$ so that addition of $H_p$ to $H_T$ does not alter the field equations for $\Pi_{\psi}$ and $\Pi_{\bar{\psi}}$.
\section{Pilot-Wave Approach in Scalar Quantum Field Theory}
In this section I consider a massive scalar field and find a Hamiltonian that gives Schrodinger's equation plus the guidance equation for the field (see \cite{Bohm4,Hol3} for discussions on the pilot-wave theory for scalar fields).

Consider a real scalar field $\phi(x^{\mu})$ with the Lagrangian
\begin{equation}
L=\frac{1}{2}\partial_{\mu}\phi\partial^{\mu}\phi-\frac{1}{2}m^2\phi^2\;.
\end{equation}
The canonical momentum is given by
\begin{equation}
\Pi=\frac{\partial\phi}{\partial t}
\end{equation}
and the Hamiltonian is given by
\begin{equation}
H=\frac{1}{2}\int\left[\Pi^2+(\nabla\phi)^2+m^2\phi^2\right]d^3x\;.
\end{equation}
Schrodinger's equation for this theory is given by
\begin{equation}
i\frac{\partial\Psi}{\partial t}=\frac{1}{2}\int\left[-\frac{\delta^2}{\delta\phi^2}+(\nabla\phi)^2+m^2\phi^2\right]d^3x\;\Psi\;.
\end{equation}
The wave function can be written as
\begin{equation}
\Psi=Re^{iS}
\end{equation}
and the equation of motion is taken to be
\begin{equation}
\Pi=\frac{\partial\phi}{\partial t}=\frac{\delta S}{\delta\phi}\;.
\end{equation}
Consider confining the field to a box of length $L$ with periodic boundary conditions. The field can be written as
\begin{equation}
\phi(\vec{x},t)=\sqrt{\frac{1}{V}}\sum_{\bf{k}}q_{\bf{k}}(t)e^{i\bf{k}\cdot\bf{x}}\;,
\end{equation}
where
\begin{equation}
{\bf{k}}=\frac{2\pi}{L}(n_x,n_y,n_z)\;,
\end{equation}
$n_x, n_y$ and $n_z$ integers, $q_{\bf{k}}^*=q_{-\bf{k}}$ and $p_{\bf{k}}^*=p_{-\bf{k}}$.
Schrodinger's equation is given by
\begin{equation}
i\frac{\partial\Psi}{\partial t}=\sum_{{\bf{k}}/2}\left[-\frac{\partial^2}{\partial q_{\bf{k}}\partial q^*_{\bf{k}}}
+(k^2+m^2)q_{\bf{k}}q^*_{\bf{k}}\right]\Psi\;,
\label{SE}
\end{equation}
where  $\Psi=\Psi(q_{\bf{k}},q^*_{\bf{k}},t)$ and ${\bf{k}}/2$ indicates that the sum is carried out over half of the values of ${\bf{k}}$.

A Lagrangian that produces (\ref{SE}) is
\begin{equation}
L=\frac{1}{2}i\left[\Psi^*\frac{\partial\Psi}{\partial t}-\Psi\frac{\partial\Psi^*}{\partial t}\right]-
\sum_{{\bf{k}}/2}\left[\frac{\partial{\Psi}}{{\partial q_{\bf{k}}}}\frac{\partial{\Psi^*}}{{\partial q^*_{\bf{k}}}}+
(k^2+m^2)q_{\bf{k}}q^*_{\bf{k}}\Psi\Psi^*\right]\;.
\end{equation}
Following the same procedure that was discussed in section 2 leads to the
constraints
\begin{equation}
\Pi_{\Psi}\approx\frac{1}{2}i\hbar\Psi^*\hskip 0.4in \hskip 0.4in \Pi_{\Psi^*}\approx-\frac{1}{2}i\hbar\Psi
\end{equation}
and to the total Hamiltonian
\begin{equation}
H_T=i\sum_{{\bf{k}}/2}\int\left\{\Pi_{\Psi}\left[\frac{\partial^2\Psi}{\partial q_{\bf{k}}\partial q^*_{\bf{k}}}-(k^2+m^2)q_{\bf{k}}q^*_{\bf{k}}\Psi\right]-
\Pi_{\Psi^*}\left[\frac{\partial^2\Psi^*}{\partial q_{\bf{k}}\partial q^*_{\bf{k}}}-(k^2+m^2)q_{\bf{k}}q^*_{\bf{k}}\Psi^*\right]\right\}dqdq^*\;,
\end{equation}
where $dq=\Pi_{{\bf{k/2}}}dq_{{\bf{k}}}$ and $dq^*=\Pi_{{\bf{k/2}}}dq^*_{{\bf{k}}}$. This Hamiltonian gives the correct equations of motion assuming Gauss' theorem (and the vanishing of surface terms) in infinite dimensional spaces.
The guidance equations are given by
\begin{equation}
\dot{q}_{\bf{k}}=\frac{\partial S}{\partial q^*_{\bf{k}}}\;\;\;\;\;\;\;\;\;\;\;\; and \;\;\;\;\;\;\;\;\;\;\;
\dot{q}^*_{\bf{k}}=\frac{\partial S}{\partial q_{\bf{k}}}\;.
\end{equation}
This equation of motion follows from the Hamiltonian
\begin{equation}
H_p=\sum_{{\bf{k}}/2}\left[p_{\bf{k}}\frac{\partial S}{\partial q^*_{\bf{k}}}
+p^*_{\bf{k}}\frac{\partial S}{\partial q_{\bf{k}}}\right]\;,
\end{equation}
where
\begin{equation}
\{q_{\bf{k}},p_{\bf{l}}\}=\{q^*_{\bf{k}},p^*_{\bf{l}}\}=\delta_{\bf{kl}}
\end{equation}
and all other Poisson brackets vanish.
The equations of motion of the particle and field follow from the Hamiltonian
\begin{equation}
H=H_T+H_p
\end{equation}
plus the constraints $p_{\bf{k}}\approx 0$ and $p^*_{\bf{k}}\approx 0$.
\section{Conclusion}
In this paper I showed that the Hamiltonian
\begin{equation}
H=\frac{i}{\hbar}\int\left[\Pi_{\psi}\left(\sum_{k=1}^N\frac{\hbar^2}{2m_k}\nabla^2_k\psi-V\psi\right)-
\Pi_{\psi^*}\left(\sum_{k=1}^N\frac{\hbar^2}{2m_k}\nabla^2_k\psi^*-V\psi^*\right)\right]d^3x_1...d^3x_N
+\sum_{k=1}^N\vec{p}_k\cdot\vec{\nabla}_kS
\end{equation}
plus the constraints $\vec{p}_k\approx 0$ gives Schrodinger's equation for a collection of non-relativistic particle in a potential and the guidance equation
\begin{equation}
\frac{d\vec{X}_k}{dt}=\vec{\nabla}_kS(\vec{X},t)\;,
\end{equation}
where $\vec{X}_k$ is the position of the $k^{th}$ particle.
It is important to note that the canonical momenta $\vec{p}_k$ are not related to $\vec{v}_k$ and it is consistent to impose the constraints
$\vec{p}_k\approx 0$.

I then showed that
\begin{equation}
H=i\int\left\{\Pi_{\psi}\gamma^0\left[i\gamma^k\partial_k\psi-eA_{\mu}\gamma^{\mu}\psi-m\psi\right]+
\left[i(\partial_k\bar{\psi})\gamma^k+e\bar{\psi}A_{\mu}\gamma^{\mu}+m\bar{\psi}\right]\gamma^0\Pi_{\bar{\psi}}\right\}d^3x+p_{\mu}J^{\mu},
\end{equation}
where
\begin{equation}
J^{\mu}=\frac{\bar{\psi}\gamma^{\mu}\psi}{\sqrt{a^2+b^2}}\;,
\end{equation}
\begin{equation}
a=\bar{\psi}\psi\;\;\;\;\;\;\;\; and \;\;\;\;\;\;\; b=i\bar{\psi}\gamma^5\psi
\end{equation}
plus the constraint $p_{\mu}\approx 0$ gives the Dirac equation and the guidance equation
\begin{equation}
\frac{dX^{\mu}}{d\tau}=J^{\mu}\;.
\end{equation}

Lastly I considered the massive Klein-Gordon equation as a quantum field (not as a single particle wave equation). After decomposing the field into normal modes
\begin{equation}
\phi(\vec{x},t)=\sqrt{\frac{1}{V}}\sum_{\bf{k}}q_{\bf{k}}(t)e^{i\bf{k}\cdot\bf{x}}
\end{equation}
I showed that the Hamiltonian
\begin{equation}
\begin{array}{cc}
                          H=i\sum_{{\bf{k}}/2}\int\left\{\Pi_{\Psi}\left[\frac{\partial^2\Psi}{\partial q_{\bf{k}}\partial q^*_{\bf{k}}}-(k^2+m^2)q_{\bf{k}}q^*_{\bf{k}}\Psi\right]-
\Pi_{\Psi^*}\left[\frac{\partial^2\Psi^*}{\partial q_{\bf{k}}\partial q^*_{\bf{k}}}-(k^2+m^2)q_{\bf{k}}q^*_{\bf{k}}\Psi^*\right]\right\}dqdq^*& \\
                            & \\
                           +\sum_{{\bf{k}}/2}\left[p_{\bf{k}}\frac{\partial S}{\partial q^*_{\bf{k}}}
                           +p^*_{\bf{k}}\frac{\partial S}{\partial q_{\bf{k}}}\right]\;,
                          \end{array}\hspace {-0.12in}
\end{equation}
where $dq=\Pi_{{\bf{k/2}}}dq_{{\bf{k}}}$ and $dq^*=\Pi_{{\bf{k/2}}}dq^*_{{\bf{k}}}$,
plus the constraints $p_{\bf{k}}\approx 0$ and $p^*_{\bf{k}}\approx 0$ gives Schrodinger's equation and the guidance equations
\begin{equation}
\dot{q}_{\bf{k}}=\frac{\partial S}{\partial q^*_{\bf{k}}}\;\;\;\;\;\;\;\;\;\;\;\; and \;\;\;\;\;\;\;\;\;\;\;
\dot{q}^*_{\bf{k}}=\frac{\partial S}{\partial q_{\bf{k}}}\;.
\end{equation}
\section*{Acknowledgements}
This research was supported by the  Natural Sciences and Engineering Research
Council of Canada.

\end{document}